\documentclass[a4paper, fontsize=11pt]{scrartcl}	
\usepackage[margin=2cm]{geometry}
 
\usepackage[T1]{fontenc}
\usepackage{helvet}

\usepackage[english]{babel}															\usepackage{amsmath,amsfonts,amsthm,amssymb}

\usepackage{multicol}						
\usepackage[pdftex]{graphicx}					
		
\usepackage{url}
\usepackage{wrapfig}
\usepackage{subcaption}
\usepackage{booktabs}
\usepackage{siunitx}
\usepackage{enumitem}
\usepackage[normalem]{ulem}

\usepackage{pdfpages}
\usepackage{scrlayer-scrpage}
\ihead{}     
\ohead{\thepage}  

\ifoot{}     
\ofoot{}     

\newcommand{\themecol}[1]{{\color{yaleblue}{#1}}}

\usepackage[numbers,sort&compress]{natbib}

	\definecolor{light-blue}{rgb}{0.8,0.85,1}
	\definecolor{airforceblue}{rgb}{0.36, 0.54, 0.66}
	\definecolor{azure}{rgb}{0.0, 0.5, 1.0}
	\definecolor{bleudefrance}{rgb}{0.19, 0.55, 0.91}
	\definecolor{blue(munsell)}{rgb}{0.0, 0.5, 0.69}
	\definecolor{darkmidnightblue}{rgb}{0.0, 0.2, 0.4}
	\definecolor{steelblue}{rgb}{0.27, 0.51, 0.71}
	\definecolor{tealblue}{rgb}{0.21, 0.46, 0.53}
	\definecolor{yaleblue}{rgb}{0.06, 0.3, 0.57}
	\definecolor{applered}{rgb}{0.89, 0.02, 0.17}

\newcommand{\mysectionstyle}{\normalfont\large\bfseries\color{yaleblue}}

\setkomafont{section}{\mysectionstyle}
\setkomafont{subsection}{\mysectionstyle}
\setkomafont{subsubsection}{\mysectionstyle}
\setkomafont{paragraph}{\mysectionstyle}
\setkomafont{subparagraph}{\mysectionstyle}

\usepackage{setspace}

\usepackage[unicode, implicit]{hyperref}
\usepackage{orcidlink}
\newcommand{\authorsmetadata}{%
         Malcolm K. Druett \orcidlink{0000-0001-9845-266X}, 
         Graham S. Kerr \orcidlink{0000-0001-5316-914X}, 
         Joel C. Allred \orcidlink{0000-0003-4227-6809}, 
         Philippa K. Browning \orcidlink{0000-0002-7089-5562}, 
         Giulio Del Zanna \orcidlink{0000-0002-4125-0204}, 
         Jaroslav Dud\'ik \orcidlink{0000-0003-1308-7427}, 
         Robertus Erd\'elyi \orcidlink{0000-0003-3439-4127}, 
         Andrzej Fludra \orcidlink{0000-0002-6093-7861}, 
         David R. Graham, 
         Laura A. Hayes \orcidlink{0000-0002-6835-2390}, 
         Sarah A. Matthews \orcidlink{0000-0001-9346-8179}, 
         James A. McLaughlin \orcidlink{0000-0002-7863-624X}, 
         Christopher M. J. Osborne \orcidlink{0000-0002-2299-2800}, 
         Alex G. M. Pietrow \orcidlink{0000-0002-0484-7634}, 
         Vanessa Polito \orcidlink{0000-0002-4980-7126}, 
         Alex J. B. Russell \orcidlink{0000-0001-5690-2351}, 
         Peter F. Wyper \orcidlink{0000-0002-6442-7818}
}
\hypersetup{
    unicode=true,          
    linktocpage=true,
    pdftoolbar=true,        
    pdfmenubar=true,        
    pdffitwindow=true,     
    pdfstartview={FitH},    
    pdfnewwindow=true,      
    colorlinks=true,       
    linkcolor=yaleblue,          
    citecolor=yaleblue,        
    filecolor=yaleblue,      
    urlcolor=yaleblue, 
    pdfauthor={\authorsmetadata}
}

\usepackage{ragged2e}

\newcommand{\horrule}[1]{\rule{\linewidth}{#1}}

\title{ \normalfont 							
		\vspace{-1.025in} 	
		{\color{yaleblue}\horrule{2pt}} \\
		\huge	
		{\color{yaleblue}{{{\textbf{Understanding Solar Flares and Energetic Events: Open questions, observational requirements, and instrumental needs over the coming decade}}}}}\\ 
        
		\Large
		{\color{black}A white Paper in response to \textbf{\textsl{Space Frontiers 2035}}}\\
        {\color{black}Primary Thematic Area: \textbf{\textsl{Heliophysics}}}\\
      
		\vspace{-0.10in}
		{\color{yaleblue}\horrule{2pt}}\newline
		
            \Large
    		\vspace{-0.30in}
            \textbf{Malcolm K. Druett$^{1}$ \& Graham S. Kerr$^{2}$} \\
            \small 
            Joel C. Allred$^{3}$,
            Philippa Browning$^{4}$,
            Giulio Del Zanna$^{5}$,
            Jaroslav Dud\'ik$^{6}$,
            Robertus Erd\'elyi$^{1}$,
            Andrzej Fludra$^{7}$,
            David R. Graham$^{8}$,
            Hamish Reid$^{9}$,
            Laura A. Hayes$^{10}$,
            Sarah A. Matthews$^{9}$,
            James McLaughlin$^{11}$,
            Christopher Osborne$^{2}$,
            Alexander G. M. Pietrow$^{12}$,
            Vanessa Polito$^{13}$,
            Alexander J. B. Russell$^{14}$,
            Peter F. Wyper$^{15}$,

                    A list of signatories is appended at the end of the manuscript.\\
      	    \textsl{(1) University of Sheffield, UK  
                   (2) University of Glasgow, UK
                   (3) NASA Goddard Space Flight Center, USA\\
                   (4) University of Manchester, UK
                   (5) University of Cambridge, UK
                   (6) Czech Academy of Sciences, Czech Republic
                   (7) Rutherford Appleton Laboratory, UK
                   (8) Bay Area Environmental Research Institute, USA\\
                   (9) Mullard Space Science Laboratory, University College London, UK
                   (10) Dublin Institute for
Advanced Studies, Ireland
                   (11) Northumbria University, UK
                   (12) Leibniz-Institut f\"{u}r Astrophysik Potsdam (AIP), Germany
                   (13) Lockheed Martin Solar and Astrophysics Laboratory, USA
                   (14) University of St. Andrews, UK\\
                   (15) Durham University, UK.
                   }

        \begin{figure*}[h!]
            \centering
            \includegraphics[width=0.90\textwidth, trim = 0cm 0.cm 0cm 0.cm]{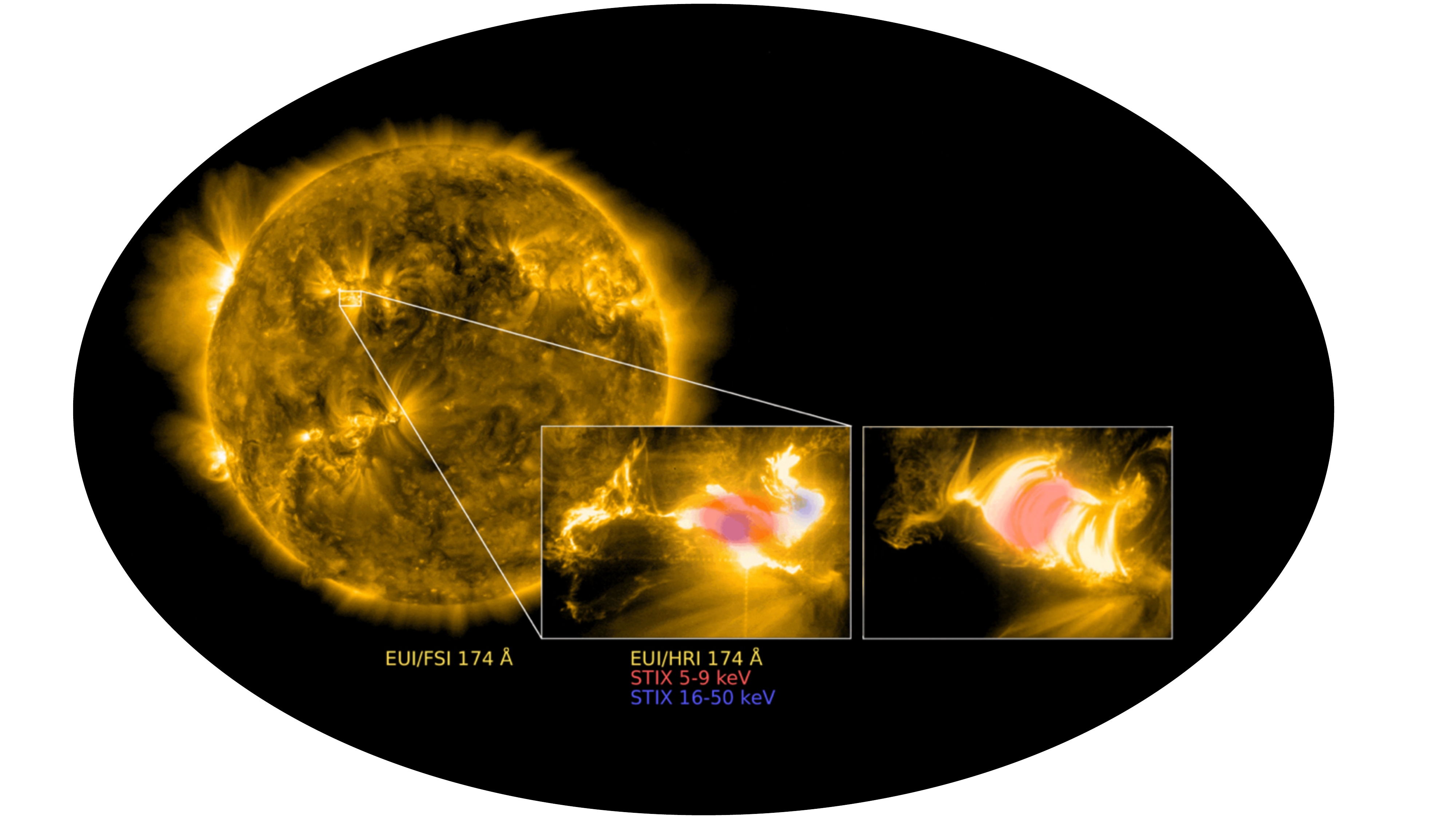}
            \caption{\textbf{A solar flare observed on 2nd March 2022 by Solar Orbiter's Extreme Ultraviolet Imager (EUI) and Spectrometer Telescope for Imaging X-rays (STIX). The insets are EUI's High-Resolution Imager (HRI), with the left being impulsive phase and the right the gradual phase.} Adapted from \url{https://www.esa.int/ESA_Multimedia/Images/2022/05/Solar_flare_2_March2}. \copyright ESA, Image credit: ESA \& NASA/Solar Orbiter/EUI \& STIX Teams}
            \label{fig:flareoverview}
        \end{figure*}
        
		}

\author{\vspace{-2em}}

\date{}

\begin{document}

\begin{spacing}{1.25}
\maketitle
\end{spacing}

\thispagestyle{empty}

 \newpage
\setcounter{page}{1}

\vspace{-1in}
\section{Executive Summary}
\vspace{-0.2in}
Solar flares are the largest energy-release events in the Solar System, allowing us to study fundamental physical phenomena under extreme conditions. Those include magnetic reconnection, particle acceleration, radiation transport, and various plasma physics processes, all of which occur throughout the heliosphere and rest of the Universe. Flares and eruptive events are also components of geo-effective space weather. Their  impacts from a space weather perspective are numerous, such as harm to satellites, disruption to GPS, communications and power systems, and impacts on passenger air travel. A comprehensive understanding of solar flares is therefore not just a compelling science problem, but also important for national security and infrastructure. This white paper (WP) addresses critical open science questions related to solar flares. 
Key observations and capabilities required to make significant advancements over the coming decade are identified. The UK has a robust and vibrant solar flare community. We are key partners in international collaborations, and also provide instrumentation for existing and upcoming ESA/NASA/JAXA space missions. Continuing this effort over the coming decade is vital to maintain UK leadership in this field, achieve Solar System Advisory Panel roadmap goals, and to work under the UKSA's National Space Strategy Pillars. Several complementary WPs have been submitted that discuss instruments or concepts that would directly address the observational requirements we describe (including: SPARK, solar optimised IFUs, Solar-C/EUVST, \& OSIRIS), as well as required numerical modelling efforts and infrastructure.

\vspace{-0.2in}
\section{Scientific motivation and objectives}
\vspace{-0.2in}

A tremendous amount of energy is stored in the Sun's magnetic field, which is released following magnetic reconnection \cite{2002A&ARv..10..313P}. The largest such events lead to solar flares (and coronal mass ejections; CMEs), during which up to $10^{32}$~erg of magnetic energy in the corona is rapidly transformed into plasma heating, particle acceleration, the generation of magnetohydrodynamic waves, and kinetic energy. Significant heating results, both in the corona but also in the lower solar atmosphere.  
Energy is transported from the coronal release site, principally along magnetic field lines (with opened field and CMEs allowing for some to escape out into the heliosphere), and a large fraction travels sunward and deposited in the lower lying chromosphere and transition region. There is evidence that some of this energy even reaches the photosphere \cite{2024ApJ...960...80S}. The primary means by which flare energy is transported through the solar atmosphere is thought to be by accelerated non-thermal electrons, owing to the unambiguous evidence of their presence in flare footpoints via hard X-ray observations \cite{2011SSRv..159..107H}, although thermal conduction, waves and ion beams likely also contribute \citep[e.g.][]{2022FrASS...960856K,2023FrASS...960862K,Lorincik2025b}.

Following flare energy release, plasma heating results in broadband enhancement to the Sun's radiation. In the (extreme-)UV, optical, and infrared this takes the appearance of `ribbons' (large scale, dynamic structures), and hard X-rays (and even $\gamma$-rays) appear as kernel-like compact sources \cite{2011SSRv..159...19F,2011SSRv..159..107H,2022MNRAS.516.6066O,2025A&A...694A..58B}. Pressure imbalance results in chromospheric ablation (commonly referred to as `evaporation') of material into the corona, where the resulting density increases boost the emission measure. Flare loops thus shine brightly in EUV and soft X-rays. With the advent of very high spatial, spectral and temporal resolution data in recent years we have also identified substantial sub-structure within flare ribbons \citep[e.g.][]{2023ApJ...944..104P, 2024A&A...685A.137P, 2025A&A...693A...8T}. What drives that substructure is not fully understood, but it is undoubtedly a critical window on the energy release, energy transport, and energy dissipation processes. 

Even with the substantial advances of recent decades, owing in large part to UK advances in the field, fundamental questions about solar flares remain unresolved:

\vspace{-0.075in}
\begin{enumerate}
    \vspace{-0.1in}
    \themecol{\item What triggers magnetic reconnection in flares, and can we predict flares before they happen?}
    Are flares initiated by ideal MHD instabilities such as torus and kink instabilities that consequentially form a current sheet, or driven by a slow photospheric or coronal process such as flux cancellation and local resistivity enhancements that can form a current sheet (which could then drive fast reconnection \citep{2023FrP....1198194K,2025SSRv..221...27D})? 
    What process generates quasi-periodic pulsations (QPPs) and can QPPs provide direct constraints on energy release and timing
\citep{2016SoPh..291.3143V,2018SSRv..214...45M,2022ApJ...925..195K, Ashfield2025,Schiavo_2025}? 
    What sets the critical thresholds of metrics for flare onsets, can directly observed or associated quantities be used to predict flares, and are these quantities causal or symptomatic of flare onset \citep{2012ApJ...747L..41B,2020ApJ...896..119K, 2021MNRAS.501.1273H, 2021ApJ...922..137W, 2024A&A...691A.119K, 2025SoPh..300....2H}?

    \vspace{-0.1in}
    \themecol{\item How, and where, are particles (electrons and ions) accelerated in flares?} 
    Copious evidence exists that energetic electrons (deka-hecta keV) are present in flare footpoints (and sometimes loops). Their energy spectra have been well-characterised via the hard X-ray emission they produce   \citep{2011SSRv..159..107H} (including with complex transport effects \citep{2019ApJ...871..225K,2024ApJ...974..119L,2025ApJ...987..211B}). However, fundamental questions remain \cite{2023BAAS...55c.206K}. The dominant acceleration mechanism(s) are not known, nor even where particles are accelerated. Possibilities include merging of magnetic islands/plasmoids in current sheets \cite{2021PhRvL.126m5101A}, acceleration in termination shocks \cite{2015Sci...350.1238C,2018ApJ...865..161P}, DC electric fields, turbulent acceleration including from fragmented current sheets \cite{2012SSRv..173..535P, 2019PPCF...61a4020V, Polito2018b, Ashfield2024}, or even local acceleration in the chromosphere  \citep{2009A&A...508..993B,2008ApJ...675.1645F}. 
    The relative role of each process, and how it varies both from flare-to-flare, and within individual flares is not known. Why are some flares more efficient accelerators than others, in terms of the number of accelerated particles and the maximum energy in their distributions? What is the anisotropy of the distributions \cite{2024ApJ...964..145J}? Are electrons and protons accelerated together, or via separate processes in time and space, and how are their properties related? Indeed we do not yet have a solid handle on the energy content and distribution of energetic ions (0.01-100~MeV), despite compelling evidence that it is comparable to that of electrons.  
    
    \vspace{-0.1in}
    \themecol{\item How is energy transported and dissipated into heating in the Sun’s atmosphere, spanning the photosphere to the corona?} Flare accelerated electrons (and likely ions) stream to the lower atmosphere where they thermalise and heat the plasma. En route, though, various effects can take place that impact the shape of the non-thermal particle distributions or how they deposit their energy, including return current effects, magnetic mirroring, wave-particle interactions, and non-Maxwellian properties of the target plasma. Key questions remain. How do precipitating electrons and ions interact with each other? How spatially variable is non-thermal particle precipitation into the lower atmosphere, and does this cause the observed small-scale structure within flares  \cite{2011SSRv..159..301K, 2011SSRv..159..357Z,2023ApJ...944..104P}? Alternately, do other energy transport mechanisms play an important, or even dominant, role at times or locations within the flare? Lower atmospheric observations have implied very deep heating, below the canonical depths which notable non-thermal electron power can reach \cite{2012ApJ...753L..26M, 2015ApJ...802...19K}. This points to additional energy transport mechanisms acting alongside electron beams, or that we are not correctly accounting for features of non-thermal electron transport. Radiation, across a wide range of wavelengths, also provides a less significant but compelling (and not yet well understood) mechanism for energy transport that is not guided by the magnetic field like non-thermal particles and plasma  \cite{2017ApJ...847...48H, 2017ApJ...837..125K, 2018A&A...610A..68D,2022MNRAS.516.6066O}.
    
    \vspace{-0.1in}    
    {\themecol{\item What is the magnitude, source, and transport mechanisms of energy deposition during the gradual phase?}}
    The flare `impulsive phase' has historically been viewed as the main energy release phase of flares, typically defined by the presence of high-energy electrons. It generally lasts several minutes, and lightcurves of X-rays and flare ribbons/footpoints more generally, show steep, rapid rise times. This is followed by the `gradual phase', which as the name implies persists over longer timescales following the main energy injection period, during which the atmosphere cools. However, flare modelling has revealed rather aggressive timescales compared to observations, with cooling occurring much too fast (an order of magnitude in some cases). Modelled mass flows also subside too quickly \citep[e.g.][]{2020ApJ...900...18K}. Though suppression of thermal conduction  \cite{2018ApJ...856...27Z,2022ApJ...931...60A} and accounting for non-equilibrium and non-Maxwellian plasmas \citep{Dudik2017} can perhaps mitigate this discrepancy, it is now widely recognised that additional energy must be deposited on longer timescales than implied by hard X-ray signatures from non-thermal electrons. Indeed, this might even require energy rivaling that of the impulsive phase \cite{2018ApJ...856...27Z}. The source and transport mechanism of this missing energy is almost completely unknown. Suggestions include Alfv\'enic turbulence \cite[e.g.][]{2023ApJ...944..147A}, slow contraction of magnetic fields, and continued reconnection from post-flare structures that results in relatively slow formation and merging of plasmoids.
  
    \vspace{-0.1in}
    \themecol{\item What is the role of turbulence in flares - from particle acceleration to its impact on plasma processes?}
    Flare line widths in flare plasma sheets, the above-the-loop region, footpoints and chromospheric evaporation imply unresolved motions \citep{Polito2019, 2024GMS...285...39R} or ion temperatures several times $T_e$ \citep{Polito2015, 2025ApJ...990L..39R}. Turbulence plays an important role in flare energy transfer \cite{2017PhRvL.118o5101K,2021ApJ...923...40S}, and 3D MHD flare simulations produce turbulence in the above-the-loop region driven by flow braking, illuminating some of the processes involved \cite{2022NatAs...6..317S, Shen2023, 2023ApJ...943..106S, 2023ApJ...954L..36W,2016A&A...589A.104G}. If above-the-loop broadening signifies turbulence, particles appear to be Fermi accelerated over approximately five seconds \citep{1993ApJ...418..912L,1994ApJ...425..856L,2017PhRvL.118o5101K}, some remaining trapped in the corona \citep{2022Natur.606..674F} and others precipitating into the footpoints. Turbulence is also integral to fast reconnection \citep{1999ApJ...517..700L}, where electrons and ions are Fermi accelerated as well \citep{2021PhRvL.126m5101A,2021PhRvL.127r5101Z,2025arXiv250800119Y}. A key question at upcoming cadences is the link between turbulence and particle acceleration, requiring joint measurements of spectral lines, hard X-rays and gamma rays (c.f. the SPARK mission WP). A new question arising from \citep{2025ApJ...990L..39R} is how to disambiguate turbulence vs. hot ions? Simulations and synthetic observations will help but the definitive solution requires observational strategies such as probing line width dependencies on ion mass and/or viewing angle. Finally, acceleration may be multi-step with a chain of injections, e.g. particles may be accelerated in the reconnection layer then at a termination shock then in above-the-loop turbulence, so different flare regions need to be integrated into a complete system model.
 
    \vspace{-0.1in}    
    \themecol{\item How does the physics of flares scale from superflares, to X-class events down to nanoflares?}
    Reconnection liberates magnetic energy across a wide range of scales, from the largest solar flares to nanoflares and so-called `campfires.' Evidence of particle acceleration exists in microflares, with a subset even exhibiting signs of very efficient electron energisation \citep{Polito2023b}. There are even hints that some tiny flares  \citep[including an A0.003 class event observed by NuSTAR,][]{2024MNRAS.529..702C}, display non-thermal signatures. An open question is, to what extent does the physics of flares scale from the very big to the very small? What are the key similarities and differences in the reconnection, particle acceleration, and plasma heating that results?
\end{enumerate}

\vspace{-0.075in}
Understanding the physics and observations of flares also provides clear and direct synergies with the other areas of heliophysics (e.g. magnetic reconnection and related process that occur in the Earth's magnetosphere; see Russell et al's WP `Magnetic Reconnection') as well as in other fields. A  noteworthy example is space weather and the near-earth environment, but they are also strongly complementary with open questions regarding the flaring activity of other stars and impacts on the habitability of surrounding planetary bodies \cite{2025Natur.643..645I}. 
\begin{enumerate}
    \vspace{-0.075in}
    \item The Solar-Stellar connection
    \begin{itemize}
        \vspace{-0.075in}
        \item How "solar like" are solar like stars in terms of flaring behaviors? What can the trends we observe in the stellar populations tell us about the probabilities of extreme events such as superflares, and changes in behaviours on the Sun \cite{2012Natur.485..478M, 2008ApJ...672..659A, 2017ApJ...851...91N, 2024LRSP...21....1K, 2025MNRAS.544.1992H}?
        \vspace{-0.075in}
        \item How do we improve our understanding of the physics of flares by combining spatially resolved solar flare observations with flares observed from a diverse stellar populations (including superflares)? Can we provide a framework to constrain and understand factors in stellar flares such as: the position of the flare on the stellar disk, the sizes of flaring areas, the identification and characterisation of CMEs, and flare energetics 
         \cite{2024A&A...682A..46P, 2024MNRAS.528.2562S, 2024ApJ...974L..13O, 2025A&A...700A.275D}.
    \end{itemize}
    \vspace{-0.1in}
    \item Driving space weather
    \begin{itemize}
        \vspace{-0.075in}
        \item Can CME and Solar Energetic Particle (SEP) properties be inferred from flare observations (including X-ray and $\gamma$-rays), which flare parameters (thermal, non-thermal, magnetic) best constrain their evolution \cite{2018ApJ...865..161P}, and how are flare accelerated particles and SEPs related? 
        \vspace{-0.075in}
        \item Do flare properties provide insight into their subsequent impacts on near Earth environment, and thereby help us to protect assets and infrastructure \cite{2014ApJ...793...70M, 2015SoPh..290.3573S}?
        \vspace{-0.075in}
        \item Can we use spectral and imaging observations from solar flares to inform or develop flare forecasting, for inferring the build of up turbulence prior to a flare \cite{2021ApJ...922..137W}.
        Including spectral information in the research-to-operations-to-research pipeline is likely fruitful \cite{2016ApJ...827..101K, 2019ApJ...873..128O}.
    \end{itemize}
    \vspace{-0.1in}
    \item Habitability and exoplanets
    \begin{itemize}
        \vspace{-0.075in}
        \item Which flare properties (energy, frequency, emissions, spectral distribution) most influence exoplanet atmospheres and photochemistry, and can these properties be inferred using techniques such as EUV spectroscopy and imaging or x-ray data? \cite{2016ApJ...829...23D, 2023MNRAS.518.2472R}
        \vspace{-0.075in}
        \item How do flare radiation and particle fluxes impact the habitability of orbiting planets, through influences such as atmospheric retention and surface condition impacts \cite{2010AsBio..10..751S, 2019AsBio..19...64T}?
    \end{itemize}
\end{enumerate}

\vspace{-0.35in}
\section{Strategic context}
\vspace{-0.2in}
Flares and solar eruptive events drive space weather and, as noted above, comprehensive understanding of the physical processes at play are crucial in our efforts to predict their occurrence, magnitude, and therefore impact. This has tangible societal impacts through impacts of space weather on critical infrastructure, communications and national security matters. The UK's international leadership in flare science also has real links to the national interest, and space economy, as evidenced by relevance to various aspects of the Solar System Advisory Panel (SSAP) roadmap:\\
\themecol{S2.2} What role do small- and large-scale flows at the photosphere play in the build-up of energy?\\
\themecol{S2.3} What triggers solar eruptions, and can we predict them?\\
\themecol{S2.4} How is magnetic energy transferred into other forms of energy?\\
\themecol{S2.6} How do non-equilibrium processes affect the plasma modelling and diagnostics?\\
\themecol{S3.2} What physical mechanisms produce geoeffective events, and how can we better forecast them?\\
\themecol{P3.3} What are the requirements for, and bounds of, habitability in the Solar System and elsewhere?\\ 
\themecol{SP3} What is the correct, self-consistent model that describes space plasma turbulence?\\
\themecol{SP4} How, and on what space/timescales, does magnetic reconnection convert energy and change space plasma magnetic field topology?\\
\themecol{SP7} How do space plasma processes efficiently accelerate different particle species?\\
\themecol{SP8} How completely can we understand and interpret the sources of electromagnetic emission?

In the SSAP roadmap, space-based and other key instrumentation are identified to address the Themes: ``1: Solar Variability and its Impact on Us'', ``2: Planets and Life'', and ``3: Space Plasma Processes''. Solar instruments explicitly highlighted include Solar Orbiter and Solar-C/EUVST. The SPARK concept is also noted as highly relevant for Theme 1. Ensuring the the UK remains at the forefront of international collaborations in solar flares, including our development of next-generation EUV/UV instrumentation, is also consistent with the UKSA's  National Space Strategy Pillars

\vspace{-0.2in}
\section{Proposed approach and key observational requirements}
\vspace{-0.2in}

Addressing the science questions identified in Section 2 demands coordinated observations that characterise both the thermodynamic and magnetic field evolution within the Sun's atmosphere, at the timescales on which flares operate: sub-second-to-several-second cadences during the flare impulsive phase. At the same time, recent observations have revealed remarkable rapidly-evolving small-scale features within flares, which undoubtedly impart information about the physical processes at play \cite{2023ApJ...944..104P,2024A&A...685A.137P,2025A&A...693A...8T,2025ApJ...990L...3T, 2025NatAs...9...45L, Ashfield2025}. Fine-scale imaging and spectral observations of the lower atmosphere are necessary windows on energy release, transport and particle acceleration \cite{2015ApJ...810....4B,2021ApJ...922..117F,2022ApJ...926..218N,2024ApJ...970...21K,2025ApJ...993...31D}, which are challenging to observe directly in the corona. Here we discuss required observations (with loose thresholds in \textbf{bold text}), driven by open science questions, best achieved by space-borne instruments\footnote{Though coordinated ground-based observations covering optical/IR and radio emission would of course be extremely valuable, we restrict ourselves here to observations that are better made in space}:\\

\vspace{-0.15in}
\themecol{\noindent\textbf{Hard X-rays}} (signatures of accelerated electrons) should be observed with sufficient spatial resolution to isolate footpoints from other flare structures. Sufficient temporal resolution is needed to characterise elementary bursts (down to $\sim$1~s or less \cite[e.g.][]{1983ApJ...265L..99K,2006SoPh..236..293Q,2023A&A...671A..79C}), and to identify acceleration timescales \citep[e.g.][]{1998ApJ...495L..67T,2013ApJ...777...33C,2018ApJ...867...84G}. A high dynamic range is essential to co-image much fainter sources within the loops and the bright footpoints. The energy range should be low enough to allow measurement of the low-energy cutoff, while high enough to characterise the most energetic particles and spectral features. \textbf{$<1$s spectra / lightcurves, $\sim3-10^{\prime\prime}$ \& 1-10s imaging, $E\sim[1-300]$~keV \& dynamic range $>100:1$.} \\

\vspace{-0.15in}
\themecol{\noindent\textbf{$\gamma$-rays}} are required to constrain properties of flare-accelerated ions. Currently we must rely on relatively rare observations \cite{2025A&A...694A..58B}, or on inferences from models. Those models are either direct particle acceleration simulations \cite{2024ApJ...974...74Y} or indirect by studying the impact of including estimated non-thermal proton distributions in radiation hydrodynamic simulations. Imaging spectroscopy of numerous nuclear $\gamma$-ray line emission $>1$~MeV and secondary emission (neutrons and positrons produced by ion collisions produce the 0.511~MeV and 2.223~MeV lines) can provide crucial information about the energy content of flare accelerated ions, and about the ambient plasma where they are produced.  \textbf{0.5-100~MeV with energy resolution of a few keV (to measure line widths), imaging with $<12^{\prime\prime}$ angular resolution (to locate source positions/morphology relative to X-rays and flare ribbons), cadence < $60$~s (to capture evolution within the impulsive phase), and sensitivity much greater than that of RHESSI (to capture signatures from M \& X flares). } \\

\vspace{-0.15in}
\themecol{\noindent\textbf{Soft X-rays}}  Addressing questions related to the hottest flare plasma, including the role of turbulence and evolution of elemental abundances, requires spectra from ions with different masses at $T_e>10$~MK, with spatial information (i.e. not disk-integrated) and good $T_e$ diagnostics. A soft X-ray spectrograph covering hydrogen-like and helium-like ions, \cite{2021FrASS...8...33D}, building on heritage like Yohkoh BCS but with spatial resolution would be invaluable.  \textbf{$\sim0.1-15$~keV with spectrally resolved high-temperature lines, $\sim1-5^{\prime\prime}$, and cadence of $<5-10$~s. For more details see the WP by Del Zanna et al. `Plasma diagnostics of solar flares with X-ray spectroscopy'}\\

\vspace{-0.15in}
\themecol{\noindent\textbf{UV/EUV Imaging}} of flares must be optimised to avoid saturation and blooming that has impacted the majority of large events in the SDO/AIA era. IRIS' Slit Jaw Imager in the UV and Solar Orbiter's Extreme Ultraviolet Imager have demonstrated the the wealth of information obtained from short exposure, high-cadence, images of both flare footpoints and flare loops \cite{2025NatAs...9...45L,ryanoverview,2024A&A...692A.176C}. A flare-optimised very high-resolution imager sampling photosphere through corona and hot flare plasma could identify fundamental flare structures and dynamics (footpoint area, ribbon width, pre-, active-, and post-reconnection loop structures and slipping reconnection \citep[e.g.][]{Dudik2014,2025NatAs...9...45L}), which have yet to be fully resolved. These are needed to constrain energetics, place lower-resolution information from energetic particles in context, and to probe variability in space and time of flare processes (including characterisation and localisation of quasi-periodic pulsations, QPPs, and other oscillatory sources or wave phenomena, \citep[e.g.][]{Ashfield2025}).  \textbf{Multi-temperature coverage, with several passbands that isolate chromosphere, corona and flare plasma,  $\sim0.2-0.3^{\prime\prime}$, $\sim0.1-2$~s cadence.} \\

\vspace{-0.15in}
\themecol{\noindent\textbf{UV/EUV Spectroscopy}} provides plasma diagnostics that allow us to characterise the evolution of electron temperature, density, coronal and chromospheric magnetic fields\footnote{Both chromospheric and coronal magnetic field observations remain elusive, but are essential in so many areas of solar physics, including flares. Measuring coronal magnetic field strengths in active regions before, during and after flares can be achieved using EUV magnetically-induced transitions \citep[MIT; a challenging but achievable diagnostic e.g.][]{2022ApJ...938...60M}. Chromospheric fields could be obtained via NUV or IR spectropolarimetry.}, mass flows, abundances, turbulence, and other properties such as departures from ionisation equilibrium. Depending on the diagnostic and science goal, the driving resolution and cadence differs. A broad temperature and spatial coverage is essential to fully characterise the flaring region. Though immensely successful to date, single-slit spectrometers such as EIS \cite{2007SoPh..243...19C}, IRIS \cite{2014SoPh..289.2733D} and SPICE \cite{2020A&A...642A..14S} have demonstrated that next-generation spectrographs must be able to capture larger fields of view either simultaneously (an IFU-type instrument) or via rapid ($\sim$<40s~s) scans. Each approach targets different science. A single-slit could allow spatial resolution to characterise the plasma on fine (sub-arcsecond) scales but with  reduced cadence owing to the need to scan the region (e.g. EUVST which we fully endorse \cite{2019SPIE11118E..07S}). Spectra from every pixel in a 2D field of view (e.g an integral field spectrometer) would likely not yet achieve sub-arcsecond resolution but could provide full plasma diagnostics on fast timescales at scales closer to 1$^{\prime\prime}$ (sub-second for strong lines, or 1-3~s for weaker lines). This would allow the general spatial variability and response to be captured simultaneously, and spatial connections across temperature identified. 
\textbf{[$\sim0.2-0.4^{\prime\prime}$ + sub-second sit-and-stare / <40~s scans of portion of active region (AR) / < 2 min scans of full AR] \textsl{or} [$\sim1^{\prime\prime}$ + 0.5-3~s of large portion of AR], $10-50$m\AA\ spectral resolution, $\log T$ = 4-7[K] coverage including many diagnostically useful lines}.\\

 \vspace{-0.15in}
\themecol{\noindent \textbf{Numerical Modelling}} is also essential in making substantial progress in understanding solar flares, and heliophysics generally. Parallel efforts that advance our numerical modelling capabilities are required alongside state-of-the-art observations. This is (1) to help interpret observations and develop new diagnostics, (2) to identify where our models fail to stand up to observational scrutiny, and (3) to help hone future instrumental needs. This includes resources that underpin much of our research and analysis, such as atomic physic databases (e.g. the UK-led CHIANTI project) and radiation transfer tools. Indeed both numerical modelling and other resources are broadly required across space physics. See the Space Frontier submission `Numerical Modelling for Solar and Space Applications'.\\

\vspace{-0.15in}
\noindent We do not propose that the UK tackle \textsl{all} of these observational needs, but we do think the UK should actively contribute to international efforts to develop instrumentation to attack the flare question across the electromagnetic spectrum. The UK does have a strong heritage of UV/EUV instrumentation (e.g. EIS, SPICE, EUVST), and should continue to play a key role in development of instruments to provide flare-optimised observations. Indeed, the Spectral Imager of the Solar Atmosphere \citep[SISA;][]{2024Aeros..11..208C} is a UK-led instrument concept that forms part of the UK-led SPARK M-class mission concept that targets flare science (c.f. SPARK and Solar EUV IFUs WPs). SISA would provide 2D EUV spectra at high-cadence. A wealth of plasma diagnostics (including coronal magnetic fields) would be obtained, spanning a broad temperature range, providing vital information to understand the dynamic evolution of the whole flaring structure that is not possible from traditional single-slit spectrometers.

\vspace{-0.2in}
\section{UK leadership, capability and partnerships}
\vspace{-0.2in}
The UK Solar Physics community (UKSP) has over 150 researchers working across the breadth of development and use of instrumentation, conducting and interpreting observations, theoretical interpretations, and numerical simulations. Within this, the UK flare research community is uniquely placed to provide leadership in many of the objectives, strategies, and approaches outlined. 

We have successful longstanding mission heritage and operations, including SOHO's Coronal Diagnostic Spectrometer \cite{1995SoPh..162..233H}, Hinode's EIS \cite{2007CulhaneEIS}, and Solar Orbiter's Spectral Investigation of the Coronal Environment (SPICE) instrument \citep[][]{2020A&A...642A..14S}. The modelling of flares, including their various emissions is now well supported in the UK. Home-grown tools \cite[e.g.][]{2021ApJ...917...14O} are in routine use, with future coupled multidimensional models in development as part of the Solar Atmospheric Modelling Suite. 

The UK has a strong lineage and active research base including key solar physics centres spread across the nation (with World leading groups located in each of England, Scotland, Northern Ireland and Wales), who are well placed to leverage data from novel space-based instrumentation, combined with modelling. 

 Such work can be leveraged to provide insight to the triggering of flares through precursor brightenings associated and magnetic field evolution \citep[e.g.][]{2024A&A...685A.137P, 2024A&A...691A.119K, 2025arXiv250701169S}, the transport and impacts of particle beams \citep[e.g.][]{2023ApJ...944..104P, 2024ApJ...970...21K, 2024A&A...684A.171D} to diagnose lower atmospheric dynamics and address the energy transport and deposition in flares \citep[e.g.][]{2014ApJ...793...70M, 2016ApJ...827..101K, 2021MNRAS.507.1972O, 2022FrASS...960856K, 2023FrASS...960862K}, and the scaling laws of flares observed on the sun and other stars \citep[e.g.][]{2020MNRAS.494.3596D, 2024LRSP...21....1K, 2024MNRAS.528.2562S, 2025A&A...700A.275D}.

By supporting solar flare instrumentation and post-launch data analysis (both our own and partner agency missions), we can leverage UK expertise in high-resolution spectroscopy, space-based observations, and flare modelling, to maintain leadership in international flare research. Our role in the field is strengthened through collaboration with ESA, NASA, JAXA, and other international missions. These missions will enable discovery, advance our understanding of flares and their impacts, contribute to the UK space economy, and benefit society by helping to protect critical infrastructure and by providing opportunities for future generations of researchers, engineers, and technologists.

\newpage
\pagenumbering{gobble}
\section*{\Large List of Signatories}

\noindent Malcolm Druett \hfill University of Sheffield, UK

\noindent Graham S. Kerr \hfill University of Glasgow, UK

\hfill Catholic University of America, USA

\noindent Keshav Aggarwal \hfill Physical Research Laboratory, India

\noindent Joel C. Allred \hfill NASA/Goddard Space Flight Center, USA

\noindent
\noindent D. Shaun Bloomfield \hfill Northumbria University, UK

\noindent Mario Bisi \hfill UKRI STFC, RAL Space, UK

\noindent Gert Botha \hfill Northumbria University, UK

\noindent David H. Brooks \hfill Computational Physics Inc., USA

\noindent Philippa K. Browning \hfill University of Manchester, UK

\noindent Joel T. Dahlin \hfill University of Maryland, USA

\noindent Abhirup Datta \hfill Indian Institute of Technology Indore

\noindent Jackie Davies \hfill UKRI STFC, RAL Space, UK

\noindent Giulio Del Zanna \hfill University  of Cambridge, UK 

\hfill University of Leicester, UK 

\noindent Jaroslav Dud\'{i}k \hfill Astronomical Institute, Czech Academy of Sciences, Czechia 

\noindent Robertus Erdelyi \hfill University of Sheffield, UK

\noindent Viktor Fedun \hfill University of Sheffield, UK

\noindent Lyndsay Fletcher \hfill University of Glasgow, UK

\noindent Andrzej Fludra \hfill UKRI STFC, RAL Space, UK

\noindent David R. Graham \hfill University of Glasgow, UK

\noindent Harry Greatorex \hfill Queen's University Belfast, UK 

\noindent Lucie Green \hfill UCL-MSSL, UK

\noindent Iain G. Hannah \hfill University of Glasgow, UK

\noindent Laura A. Hayes \hfill Dublin Institute for Advanced Studies, Ireland

\noindent Andrew Hillier \hfill University of Exeter, UK

\noindent Hugh S. Hudson \hfill University of Glasgow, UK

\noindent Natasha Jeffrey \hfill Northumbria University, UK

\noindent Konstantinos Karampelas \hfill KU Leuven, Belgium

\noindent Eduard Kontar \hfill University of Glasgow, UK

\noindent Marianna Korsos \hfill University of Sheffield, UK

\noindent Anshu Kumari \hfill Physical Research Laboratory, India

\noindent David Long \hfill Dublin City University, Ireland 

\noindent Luke Majury
\hfill Queen's University Belfast, UK

\noindent Mihalis Mathioudakis \hfill Queen's University Belfast, UK 

\noindent Sarah A. Matthews
\hfill UCL-MSSL, UK

\noindent James McKevitt \hfill UCL-MSSL, UK

\noindent James A. McLaughlin \hfill Northumbria University, UK

\noindent Teodora Mihailescu \hfill INAF Osservatorio Astronomico di Roma, Italy

\noindent Ryan O. Milligan \hfill Queen's University Belfast, UK

\noindent Aaron Monson \hfill Queen's University Belfast, UK

\noindent Sargam Mulay \hfill University of Glasgow, UK

\noindent Thomas Neukirch \hfill University of St. Andrews, UK

\noindent Aisling O'Hare \hfill Queen's University Belfast, UK

\noindent Christopher M. J. Osborne \hfill University of Glasgow, UK 

\noindent Divya Paliwal \hfill Physical Research Laboratory, India

\noindent Ross Pallister \hfill Northumbria University, UK

\noindent Alexander G. M. Pietrow \hfill Leibniz-Institute for Astrophysics Potsdam (AIP), Germany

\noindent Vanessa Polito \hfill Lockheed Martin Solar and Astrophysics Laboratory, USA

\noindent Hamish A. S Reid \hfill 
UCL-MSSL, UK

\noindent Alexander J. B. Russell \hfill University of St. Andrews, UK

\noindent Daniel F. Ryan \hfill 
UCL-MSSL, UK

\noindent Luiz Schiavo \hfill Northumbria University, UK

\noindent Morgan Stores \hfill University of Minnesota, USA 

\noindent Peter F. Wyper \hfill Durham University, UK

\noindent Stephanie Yardley \hfill Northumbria University, UK

\noindent Natalia Zambrana Prado \hfill UCL-MSSL, UK

\noindent Jinge Zhang \hfill Paris Observatory PSL, France

\noindent Serhii Zharkov \hfill University of Hull, UK


\newpage
\pagenumbering{gobble}
\begin{spacing}{1.2}
	\bibliography{references}
\end{spacing}


\end{document}